# Low-Symmetry Two-Dimensional Materials for Electronic and Photonic Applications

He Tian[1], Jesse Tice[2,*], Ruixiang Fei[3], Vy Tran[3], Xiaodong Yan[1], Li Yang[3], Han Wang[1,*]

[1]Ming Hsieh Department of Electrical Engineering, University of Southern California, Los Angeles, CA 90089, USA
[2]NG Next, Northrop Grumman, 1 Space Park, Redondo Beach, CA 90278, USA
[3]Department of Physics, Washington University in St Louis, Missouri 63130, USA

*Corresponding authors: Jesse.Tice@ngc.com (J.T.), han.wang.4@usc.edu (H.W.)

**Introduction**

In the past decade, the research community has seen intense research efforts in the field of two-dimensional (2D) materials. Many of the common 2D materials, such as graphene, hexagonal boron nitride (hBN) and molybdenum disulfide ($MoS_2$), have relatively symmetrical 2D crystal lattices, resulting in mostly isotropic in-plane physical properties. The electrical, optical and phonon properties of these materials are similar along the different in-plane crystal directions of their 2D lattice. Having similar properties along different crystal orientations are favorable for many conventional electronic and photonic applications. For example, transistors built along different crystal directions on the wafer will have the same characteristics, which is beneficial for controlling the variability and uniformity of the device performance across a large wafer. On the other hand, there is a selected group of emerging 2D layered materials that possess highly asymmetrical in-plane crystal structure. Such layered materials include black phosphorus (BP) and its arsenic alloys, compounds with isoelectronic structure as BP including the monochalcogenides of group IV elements like Ge and Sn, as well as a class of low-symmetry transition metal dichalcogenide (LS-TMDC) materials such as rhenium disulfide ($ReS_2$) and rhenium diselenide ($ReSe_2$) (Figure 1, upper panel). Due to their reduced crystal symmetry, they have distinct electrical, optical, thermal and mechanical characteristics along different in-plane crystal directions. This new degree of freedom in their physical properties, together with other unique features arising from their low symmetry crystal lattice, can provide previously unexplored tunability on the characteristics of electrical, optical, thermal and mechanical devices (lower panels of Figure 1), hence opening the door to a wide range of opportunities for developing conceptually new semiconductor device applications (Figure 1, lower panel). In this review, we will discuss the recent progresses in studying the fundamental properties and novel device applications of low-symmetry 2D crystals. We will also offer our perspectives on the promising future research directions related to this emerging class of materials.

*Black Phosphorus* is a low-symmetry layered material that has recently received much attention. Similar to graphite, bulk black phosphorus is an elemental layered material and is the most stable allotrope of phosphorus due to its unique puckered orthorhombic crystal structure. Similar to the carbon atoms in graphene, each phosphorus atom in BP is connected to three adjacent phosphorus atoms to form a stable, linked ring structure, with each ring consisting of six phosphorus atoms (Figure 1, upper panel). The adjacent layer spacing is around 0.53 nm and the lattice constant in this orthorhombic system along the z-direction is 1.05 nm. The crystal structure of BP belongs to space group Cmca (a=3.31, b=10.54, c=4.36). Within each layer of BP, the phosphorus atoms stay on two separate planes with shorter in-plane P-P bonds (0.2224 nm) and longer P-P bonds (0.2244 nm) between the top and bottom atomic planes. For the top view of the BP lattice in the z-direction, the BP shows a hexagonal structure with two different bond angles (96.3° and 102.1°). The crystal lattice has armchair ridges along its x-direction and zigzag ridges along its y-direction (Figure 1, upper panel) resulting in its anisotropic in-plane electrical, optical and phonon properties. Mono- and few-layer black phosphorus also exhibit strong in-plane anisotropy as in bulk black phosphorus, which has interesting implications in its properties such as the anisotropic exciton behaviors and orthogonal thermal and electronic transport. Another important feature of BP is its tunable direct bandgap with thickness, which increases from 0.3 eV in the bulk to above 1.3 eV in the monolayer. Additionally, Kim et al. [1] demonstrated that the bandgap of BP can be tuned by the electric field due to the giant Stark effect. This 0.3 eV bandgap semiconductor can be turned into a band-inverted semimetal with potassium ion doping on its surface. Furthermore, the bandgap of BP can be reduced below 0.3 eV by alloying it with arsenic (As) to form black arsenic-phosphorus (b-$As_xP_{1-x}$) [2] (Figure 1, upper panel). The layered b-AsP, which has retained the puckered orthorhombic structure of BP, has a bulk bandgap ranging from 0.15 eV to 0.3 eV depending on the elemental composition. The material is hence attractive for exploring photonic applications in the mid-infrared (mid-IR) and long wavelength infrared (LWIR) range from 3 μm to 8 μm wavelength and beyond.

Materials with puckered orthorhombic crystal lattice also exist in the form of bi-elemental compounds. *Group IV Mono-Chalcogenides (SnSe SnS, GeSe and GeS)* possess similar crystal lattice as BP and share its low-symmetry structure and properties. The bandgap of these materials ranges from 0.9 eV (bulk) to 1.9 eV (monolayer), which are higher than that in BP and b-$As_xP_{1-x}$. The corresponding wavelength of optical response can cover the near-IR and part of the visible spectral range. Recent conductivity measurements indicated a mobility ratio up to ~4 [3] in bulk SnSe along different crystal directions, demonstrating even stronger anisotropy compared to BP due to its bi-elemental composition [4, 5]. Moreover, theoretical study predicted that bilayer and monolayer GeSe possess direct bandgap [6], which could be attractive for certain optoelectronics application in the near-IR and visible range. The most distinct features of this class of low-symmetry 2D materials may lie in their piezoelectric and thermoelectric properties that will be discussed in detail in later sections. For example, GeS and SnSe are predicted to possess giant piezoelectric coefficients (Figure 1, lower panel) and record high thermoelectric figure of merit has been experimentally demonstrated in SnSe (Figure 1, lower panel), making them attractive for developing high performance piezoelectric and thermoelectric devices.

2D materials with low level of in-plane crystal symmetry also occur in the form of transition metal dichalcogenides (TMDCs). Such materials include *ReSe$_2$* and *ReS$_2$* [7]. As shown in Figure 1, every unit cell of monolayer ReSe$_2$ contains four formula units, which includes two categories of Re atoms together with four categories of Se atoms. The Se atoms sandwich the Re atoms in the middle to form a monolayer lattice of ReSe$_2$. Unlike common TMDCs such as MoS$_2$ and WSe$_2$, which crystallize in the hexagonal (H) phases, ReSe$_2$ crystal displays a distorted C$_d$Cl$_2$-type lattice structure. Due to Peierls distortion, adjacent Re atoms are bonded in the form of zigzag four-atom clusters, which align along the direction of the lattice vector **a** to form Re chains (Figure 1). Theory has predicted that such a distorted octahedral (1T) crystal structure possesses lower energy than its hexagonal counterpart, thus being more stable. This clustering of Re atoms, contributing to the distortion of the lattice geometry and stabilization of the crystal structure, has also been discovered in ReS$_2$ crystals. The distorted 1T nature of ReSe$_2$ and ReS$_2$ has conferred both materials strong in-plane anisotropy in their optical and electronic properties.

In this review, we will discuss the synthesis, properties, and novel device applications of these three groups of low-symmetry 2D materials. The article is organized as follows. The next section will review the existing techniques for synthesizing low-symmetry 2D materials. Subsequent sections will review the recent work in studying the electronic, optical, and phonon properties of low-symmetry 2D materials and their applications in innovative electronic and photonic devices. The piezoelectric and thermoelectric properties of these materials and their prospect for device applications will then be discussed. We will conclude the article with an outlook of the future research opportunities and practical device applications that can be enabled by this unique class of 2D materials.

**Material Synthesis and Stability**

Here, we will first review the preparation methods for obtaining the various types of low-symmetry layered material bulk crystals, which can be exfoliated into atomically-thin layers as shown in Figure 2a-2c. The promising routes to directly access the thin film or monolayer forms of these materials will also be discussed (Figure 2d-2f). Additionally, techniques to enhance resilience of these materials towards heat, light, oxygen, and moisture will be reviewed (Figure 2g-2i).

*Black Phosphorus and its Arsenic Alloys:* Black phosphorus (BP) crystals were initially prepared by P. W. Bridgman [8] through the conversion of white phosphorus under high hydrostatic pressure. BP crystal preparation has been investigated by varying hydrostatic pressure and temperature from high pressures at room temperature to high temperatures at a relatively lower pressure (Figure 2e) [9]. Other techniques to produce BP crystals have been via metal catalyst assisted growth without the need for high pressure [10], BP microribbons via red phosphorus, Sn, and $I_2$ [11], and mineralizer-assisted vapor transport growth [12]. Furthermore, a plasma-assisted method (Figure 2c) demonstrated on mechanically exfoliated BP can thermally thin the thin films with nanometer precision [13]. A patent from Lucent in 2000 described a high vacuum preparation method for BP via thermal cycling of red phosphorus [14]. This disclosure arose out of the need to convert the more hazardous forms of phosphorus leftover from low pressure compound semiconductor growth into the more stable BP for subsequent disposal, yet a detailed procedure was never published in a refereed journal. The largest area of directly synthesized thin film BP prepared to date has been a 4 mm diameter disk that is 40 nm thick on a flexible substrate by the high pressure conversion in a multi-anvil cell of CVD deposited thin film red phosphorus to BP [9]. This method is more amenable to the microelectronics applications and demonstrates the first scalable technique for BP thin film preparation.

Bulk BP can also be alloyed with As via a novel mineralizer-assisted short way chemical transport reaction to create a b-$As_xP_{1-x}$ alloy with tunable compositions [2, 15]. This alloy was produced via a vapor transport method using ultrahigh purity elemental starting materials and metal halide catalysts of either tin iodide for the lower arsenic concentration alloys or lead iodide for the higher concentration arsenic alloys.

Even though BP is the most thermodynamically stable allotrope of phosphorus, it is still chemically reactive to the oxygen and moisture in the ambient conditions. The high surface area and reactivity of the exposed p-orbital electrons render BP susceptible to oxidation, which degrades the semiconducting properties. There have been several compelling studies on the degradation mechanisms of BP [16-18]. Favron et al. determined the most plausible degradation mechanism to be photo-induced oxidation [19]. This paper also compared in significant detail BP stability as a function of layer number, where thicker flakes were substantially more stable. This stability is attributed to a shift in bandgap to lower energy at higher layer number, which in turn decreases the oxygen acceptor states due to weaker overlap at the band edge. Additionally, there has been a report indicating that thinner BP flakes absorb water to nucleate oxidation at a quicker rate and thus wet much faster than thicker flakes [20]. Although the native oxide $PO_x$ of BP can be beneficial for some specific device concepts, such as in artificial synaptic device [21], most applications desire a more pristine BP surface. Numerous techniques have been proven to provide effective encapsulation of BP from the ambient environment including the use of exfoliated hBN [22], atomic layer deposition (ALD) of alumina (Figure 2h), chemical passivation (Figure 2g) [23] and dielectric capping (Figure 2i) [16].

*Group IV Monochalcogenides:* The low-symmetry group IV monochalcogenides consist of the compounds MX, where M = Ge, Sn and X = S, Se. The initial preparation of these MX compounds has occurred at very different times in history, because of the unique thermoelectric and optical properties, the bulk structures of MX have been widely fabricated and studied for thermoelectric and photovoltaic applications [23-25]. GeS crystals have been well known due to the seminal work of Winkler in 1886, who also discovered Ge [24]. Winkler reduced $GeS_2$ in a $CO_2$ environment with metallic Ge as a reducing agent. SnS crystals were prepared by the melt growth method in 1932 by Hofmann [25]. A very thorough review on the experimental synthesis methods for bulk crystals and thin films of SnS is noted, however, preparation of monolayers was not discussed [26].

Okazaki prepared crystals of both SnSe [27] and GeSe [28] in 1956 and 1958, respectively. These selenides were prepared by heating the elements above their melting point followed by slowly cooling the sample to the room temperature in order to measure their x-ray diffraction (XRD) patterns. Most notably, bulk SnSe was doped with Na and found to have the highest ZT to date [4]. These crystals were prepared by Kanatzidis et. al. from high purity elemental starting materials under high temperature thermal cycling in vacuum. Any manipulation of SnSe was handled in a glove box with low oxygen and moisture levels due to the propensity to oxidize. An X-ray photoelectron spectroscopy (XPS) study of bulk SnSe revealed that a native $SnO_2$ layer can form on the sample surface [29]. While not as many oxidation and encapsulation studies have been performed on SnSe compared to BP, the same practices that can protect BP may also be applicable to this material.

More recently, synthesis routes to the group IV monochalcogenides in their atomically-thin form have been demonstrated. SnSe nanoplatelets (Figure 2d) have been produced via vaporization of high purity SnSe powder [30]. These platelets were ~5 $\mu m^2$ and varied between a monolayer to ten layers of SnSe. Additionally, a recent publication on single layer, single crystalline, one pot synthesis of SnSe was published that utilizes $SeO_2$ and $SnCl_4$ pentahydrate in oleylamine and phenanthroline [31].

Most of the literature reports on the synthesis of group IV monochalcogenides have focused on 0-D and 1-D nanostructures with fewer reports on true 2D layered materials, and of these reports, most have focused on nanosheets with typical lateral dimensions less than 1 $\mu m$, and thickness ranging from monolayers to 20 nm. The size limitation of these nanosheets, or those from solution based preparation methods, typically render them difficult to use for scalable electronics application. A major challenge in all of these compounds lies in devising scalable vapor transport schemes to yield high quality, large area, and controllable crystalline layers. There have been several reports of vapor deposited group IV monochalcogenides in the literature. Li et al. have shown a unique boundary layer diffusion mechanism that occurs in GeS nanosheets [32]. Large SnS 2D flakes with variable thickness were produced via physical vapor deposition on mica substrates [33], however the very thin layers at ~5.5 nm were found to have weak and broad Raman signals. Brent et al. have demonstrated SnS bilayers through liquid exfoliation in NMP solvent [34]. Additionally, Ramasamy et al. [35] have investigated a number of different methods to obtain SnS including organometallic aerosol assisted CVD precursors for the production of vertically aligned SnS nanosheets, however monolayer films were never explored. This specialized CVD technique with unique precursor chemistries and pre-formed bonds of desired stoichiometry can be a promising route for scalable synthesis of 2D layered materials.

*Low-Symmetry Transition Metal Dichalcogenides:* The low-symmetry transition metal dichalcogenides consist of $ReS_2$ and $ReSe_2$. Among the TMDC class of layered materials, the rhenium chalcogenides have historically been some of the least studied. In fact, the Raman spectra of any type of pure $ReSe_2$ had not even been reported until 2014. The first crystals of $ReSe_2$ were prepared by a wet reaction synthesis intermediate $Re_2Se_7$, which was decomposed under vacuum at 325°C to yield $ReSe_2$ [36]. Additionally, the bulk crystals of these rhenium chalcogenide layered materials have been synthesized via vapor transport directly from the elements at high temperatures, where a halide carrier reportedly improved the crystallinity [37].

Similar to the other low-symmetry 2D materials, the LS-TMDC studies have been based on mechanically exfoliated samples from bulk crystals (Figure 2a), however the resulting flakes have typical lateral size in the range of few to tens of micrometers. The demand for high quality larger scale materials has spurred CVD processes, and larger area growth of $ReS_2$ atomic layers have been demonstrated from high purity elemental powders [38]. This method for synthesizing LS-TMDCs onto SiO$_2$/Si wafers can yield different morphologies including flakes and nanoribbons depending on the growth time, ranging from flakes at short growth times to ribbons at longer growth times (Figure 2f). Additionally, Keyshar et al. have synthesized ReS$_2$ monolayers via CVD at considerably low growth temperatures of 450°C using sulfur and ammonium perrhenate [39]. Fujita et. al. [40] has produced ReS$_2$ nanosheets using liquid exfoliation via lithium borohydride intercalation (Figure 2b). The process is different from that used for the synthesis of other TMDCs, which are typically liquid exfoliated using butyl lithium. The synthesis of ReSe$_2$ has been studied much less extensively than ReS$_2$, however, an abstract at the American Physical Society March Meeting 2016 regarding high quality CVD synthesis of thin film ReSe$_2$ was found during the preparation of this review [41].

In general, the research in the synthesis of low symmetry 2D materials is still in the infant stage. There are many opportunities for research innovations in developing new synthesis approaches and novel growth mechanisms with the goal of achieving high-quality large-scale materials at relatively low cost. The success in the synthesis research is not only of great importance from the perspective of fundamental materials study, but also critical for laying the foundation towards the applications of these materials in electronics, photonics, piezoelectronics and thermoelectrics, which we will discuss in detail in the following sections.

**Anisotropic Transport Properties and Novel Electronic Devices**

Low-symmetry 2D materials offer strong in-plane anisotropy in their transport properties that is typically a result of the different energy band structure along the different in-plane directions of the layered crystal lattice, leading to drastically different carrier effective mass along the different crystal directions. Here, we will review the transport properties of low-symmetry 2D materials using black phosphorus as an example material, and discuss the novel electronic device concepts that can be enabled by the transport anisotropy. Many of the same concepts can also be readily applied to other low-symmetry 2D materials.

The room temperature mobility of BP can reach above 1000 cm$^2$/V.s for both electrons and holes while the hole mobility can reach above 50,000 cm$^2$V$^{-1}$S$^{-1}$ at low temperature [42]. For both single-crystalline bulk and thin film BP samples, since the carrier effective mass along the x-direction of the crystal is about 10 times smaller than that along the y-direction, the carrier mobility along the x-direction can be twice the value along the y-direction for both holes and electrons [43]. Theoretical results also show abnormal behavior for monolayer BP [43]. The electron mobility along the x-direction is predicted to be about 14 times the values in the y-direction while the hole mobility in the x-direction is 16-38 times smaller than that along the y-direction [43]. This can be explained by the extremely small deformation potential along the y-direction. The two-dimensional electron gases (2DEG) in bulk or thin film BP created through electrostatic bias also allows some fundamental quantum phenomena to be observed. Li et al. [44] reported high mobility exceeding 3900 cm$^2$V$^{-1}$s$^{-1}$ at 1.5 K for BP on h-BN, allowing quantum oscillation to be observed in BP for the first time. Due to the anisotropic carrier effective masses, the amplitudes of the quantum oscillations vary along different crystal directions. Chen et al. [45] and Gillgren et al. [46] also observed quantum oscillation in high quality h-BN/BP/h-BN heterostructures. Most recently, in a graphite/h-BN/BP/h-BN heterostructure (Figure 3a), Li et al. [47] demonstrated hall mobility up to 6000 cm$^2$V$^{-1}$s$^{-1}$ (Figure 3b) at 4 K and observed the integer quantum Hall effect in BP for the first time. These experiments highlight the excellent electronic quality of the material, especially when protected by hBN. Another interesting feature in the electronic property of BP derives from its moderate 0.3 eV bandgap. As a result, the Fermi-level in BP can be easily shifted across its bandgap and the material can have either p-type or n-type carriers depending on the electrostatic bias. Such ambipolar conduction properties resulting from its relatively small bandgap make BP promising as an alternative candidate material to graphene for developing ambipolar electronic devices with novel functionalities such as ambipolar frequency multipliers [48, 49] and mixers [50].

The first BP field-effect transistors (FET) [51-54] were demonstrated in 2014. Li et al. [51] reported devices with on/off current ratio up to $10^5$ (Figure 3c) and respectable current saturation characteristics. These properties overcome the shortcomings of graphene FETs with low on/off current ratio and linear $I_d$-$V_d$ characteristics. Moreover, the hole mobility of BP is as high as 1000 $cm^2V^{-1}s^{-1}$ at room temperature, which is advantageous compared to commercial Si with hole mobility ~450 $cm^2V^{-1}s^{-1}$ [55] and much higher than other layered semiconductor materials such as $MoS_2$ and $WSe_2$. Another noticeable feature is that the on/off current ratio and mobility strongly depend on the thickness of BP. The on/off current ratio increases as the thickness of the FET channel decreases. The screening length in BP FETs is estimated to be around 2.9 nm [51]. Low et al. [56] also calculated the screening length in BP at a different carrier density to be around 10 nm using a nonlinear Thomas-Fermi model. On the other hand, thinner BP can be more sensitive to both the oxidation degradation and the charge impurities at the interfaces, which can suppress the carrier mobility. Koenig et al. [54] fabricated a BP FET with a 10 nm thick channel demonstrating a field-effect mobility around 300 $cm^2V^{-1}s^{-1}$. Liu et al. [52] exfoliated monolayer BP successfully, though monolayer BP FETs have been more difficult to fabricate due to its fast oxidation in ambient environment. Crystal orientation is another important parameter that can influence the BP mobility. Xia et al. [53] first revealed the hole mobility for 15 nm BP in the x- and y-direction are 1000 and 600 $cm^2V^{-1}s^{-1}$, respectively (Figure 3d). The hole mobility along the x-direction is about 1.8 times as that along the y-direction, which is close to the theoretical prediction. This prominent anisotropy is a direct result of the smaller carrier effective mass along the x-direction of the BP crystal compared to that along the y-direction.

The high mobility in thin film BP makes the material promising for high speed thin film electronics applications. In late 2014, the first BP radio frequency transistor (RF) was demonstrated by Wang et al. [57]. The device was fabricated along the x-direction of the BP crystal, which offers the highest carrier mobility. This 300 nm channel length device demonstrated a short-circuit current gain cut-off frequency of 12 GHz and maximum oscillation frequency of 20 GHz. Comparing to graphene FETs, in which the $f_{max}$ of the device is often significantly lower than its $f_T$, the improved current saturation characteristics in BP devices due to its non-zero bandgap is beneficial for achieving a higher power gain in the device, as reflected in its higher $f_{max}$ compared to $f_T$. Zhu et al. [49] further demonstrated a BP modulator, frequency doubler, and inverter on a flexible substrate, which shows the potential of BP for applications in high frequency thin film electronics.

The anisotropic transport properties also find interesting applications for BP in the field of bio-inspired electronics (Figure 1, lower panel). In a biological neural network, synapses are functional links between neurons through which the "information" flows. Synapses in biological systems vary significantly in their connection strengths. Realizing such heterogeneity in synaptic electronics is critical towards building artificial neural network with the potential for achieving the level of complexity in biological systems. Tian et al. [21] demonstrated the first BP synapse (Figure 3e), which offers intrinsic heterogeneity in its synaptic characteristics directly resulting from its low crystalline symmetry. The native oxide $PO_x$ of BP is utilized as an electron trapping layer, giving rise to the synaptic behavior through the trapping and de-trapping of electrons in response to positive and negative pulses applied to the gate. Synaptic weight change in the y-direction device is 1.75 times higher than that of the x-direction device (Figure 3f). The different synaptic characteristics of the x- and y-direction devices is explained by the different mobility change influenced by the trap scattering along different directions of the BP crystal. The trapped charges in $PO_x$ controlled by the gate pulse can suppress the higher mobility in the x-direction more significantly. Key features of biological synapses such as long-term plasticity with heterogeneity, including long-term potentiation/depression and spike-timing-dependent plasticity, are mimicked. A simple compact heterogeneous axon-multi-synapse network using BP synapse was also demonstrated with the tuning of the synaptic device behavior enabled by the anisotropy of the BP channel for emulating the complex heterogeneity in biological neural network.

**Anisotropic Optical Properties and Novel Photonic Devices**

Besides electronic devices, the anisotropy in the physical properties of low symmetry 2D materials also offers a new degree of freedom for designing optical devices (Figure 1, lower panel). Due to the low symmetry in-plane lattice of BP, its optical absorption is highly polarization-dependent. Based on the optical selection rules, the optical conductivity maximum of BP is along the x-direction of the crystal. Yale/IBM researchers [53] measured the polarization-resolved relative extinction spectra of thin film BP in the mid-infrared range, confirming its bandgap and the optical anisotropy (Figure 4a). The moderate 0.3 eV bandgap of BP thin film (5-15 nm thick) and its tunability with a transverse electric field [1] can enable infrared photodetectors over a wide spectral range from LWIR to visible. Buscema et al. [58] first reported BP photodetectors working in the visible and near-IR range up to 940 nm wavelength with a responsivity of 4.8 mA/W and a 1 ms response time. In order to separate electron-hole pairs more efficiently, BP p-n junction photodetector was later demonstrated [59]. A dual-gate BP FET was fabricated by placing BP on h-BN dielectric and modulated by a pair of split gates. By applying biases of opposite polarity to the two gates, a p-n junction is electrostatically created in the BP sample. The photo-excited carriers can be separated by the built-in electric field based on the photovoltaic effect. Yuan et al. [60] demonstrated a polarization sensitive vertical BP p-n photodetector that shows much stronger responsivity in the x-direction than in the y-direction over the near-IR range (Figure 4b) from 400 nm to about 1600 nm wavelength. Through single pixel scanning, Engel et al. [61] demonstrated that BP photodetectors can be used for high resolution imaging both in the visible and near-IR wavelength range. The responsivity is 20 mA/W at 532 nm wavelength and 5 mA/W at 1550 nm wavelength. The gain is up to 730% at 1550 nm. The gain could be explained by the shallow trapping states in BP, which can enable one incident photon to induce multiple charge carriers. The BP photodetector shows no sign of significant photocurrent decay up to kHz frequency. Youngblood et al. [62] demonstrated BP photodetector integrated with silicon photonic waveguide operating with a bandwidth exceeding 3 GHz, hence showing the potential of BP detectors for applications in the communication wavelength. The responsivity is 135 mA/W and 57 mA/W for 11.5 nm and 100 nm thick BP sample, respectively. Most recently, Guo et al. [63] demonstrated a BP mid-infrared detector operating at 3.39 μm with high internal gain, resulting in an external responsivity of 82 A/W up to kilohertz modulation frequencies. The operating wavelength of the device falls in the 3-5 μm range, which is a critical spectral window for thermal infrared technology with applications in high-resolution IR imaging that can penetrate through haze, dust, smoke or the darkest night. Comparing to incumbent technology in this wavelength range based on mercury-cadmium-telluride (MCT), the BP devices may offer room temperature operation and easy integration with silicon based read-out circuits. Furthermore, the polarization sensitivity of the BP device resulting from its anisotropic photocurrent response can enable the device to collect rich information from the scattering light about the surface roughness [64], geometry [65], or orientation [66] of the imaged objects, in addition to the light intensity. The polarization contrast techniques can also provide additional information to enhance target contrast in optically scattered environments, such as in hazy or foggy conditions [67].The optical anisotropy of BP becomes even more significant in monolayer samples with interesting exciton behaviors due to the two-dimensional quantum confinement. Tran et al. [68] performed the first-principles density functional theory (DFT-GW) simulations and found that the excitonic effect dominates the optical properties of monolayer and few-layer BP. For monolayer BP, the first absorption peak is located at 1.2 eV, which indicates that the exciton binding energy is ~800 meV. The binding energy reduces to 30 meV in bulk BP. Wang et al. [69] experimentally determined the monolayer BP optical bandgap to be 1.3 eV and the quasiparticle bandgap to be 2.2 eV. The measured binding energy is hence 900 meV, which agrees well with the theory of Tran et al. [68, 70]. The strong polarization-dependence of the PL intensity in monolayer BP is shown in Figure 4c. Other works have reported different values of exciton binding energy in BP subject to different sample environment and measurement condition. Yang et al. [71] reported trion binding energy for monolayer BP to be 100 meV. To maintain the high quality of monolayer BP and protect the material from oxidation, Li et al. [72] sandwiched the monolayer BP between h-BN and sapphire. They found a smaller binding energy of only 80 meV due to the screening by the dielectric environment. The exciton behavior in BP will continue to be an interesting research area and more work is needed to fully explain the carrier dynamics in atomically thin BP.

The anisotropic properties of BP are also interesting for building tunable plasmonic devices. Conventional plasmonic materials are noble metals (such as gold and silver) that mainly work in the visible and near-IR frequency range [73]. In the past few years, graphene has been proposed as an interesting material for building plasmonic devices due to its tunable carrier density and unique plasmonic dispersion resulting from its Dirac cone bandstructure. However, both metals and graphene have largely isotropic in-plane conductivity. Recently, Low et al. [56] predicted that the plasmon dispersion in BP varies significantly with the change in the in-plane crystal direction due to the in-plane conductivity anisotropy in the material. The plasmonic dispersion hence has a vectorial dependence on the momentum wavevector instead of a scalar dependence as in the case of metals and graphene. BP plasmonic devices hence have an additional tuning knob, i.e. the polarization of the incident light, for dynamically changing the resonance frequency, which is not previously available in metals and graphene. Theoretical calculations shown in Figure 4d suggests that by tuning the polarization of the incident light, the resonance frequency can be continuously changed from the far-IR to the mid-IR wavelength range.

**Anisotropic Phonon Properties and Raman spectroscopy**

Recent study of LS-TMDCs and black phosphorus using optical techniques such as Raman spectroscopy has also revealed a wealth of information about the phonon activities in these materials. The intra-layer phonon modes, and the inter-layer shear and breathing modes offer deep insights into the crystal structures and phonon properties in mono- to multi-layer samples.

Intra-layer Raman modes offer information of material type, crystal symmetry, phonon behaviors, and bond strength. Black phosphorus has a highly anisotropic intra-layer Raman response due to its reduced crystal symmetry [53]. As shown in Figure 5a and 5b, there are three observable intra-layer Raman peaks in typical BP flakes in addition to the inter-layer shear and breathing modes. They are the $A_g^1$, $B_{2g}$ and $A_g^2$ modes at 365, 440 and 470 cm$^{-1}$, respectively. The intensity of $A_g^1$ mode in the y-direction is two to three times stronger than that in the x-direction, therefore has been frequently utilized to label the crystal orientation of BP flakes. The intra-layer modes of $ReSe_2$ were also studied recently [74] as a function of layer thickness (Figure 5c). For TMDCs with isotropic hexagonal lattice structure such as $MoS_2$ [75], an in-plane vibration mode $E_{2g}^1$ mode and an out-of-plane vibration mode $A_{1g}$ can be detected. Due to the in-plane symmetry, the $E_{2g}$ mode is a degenerated mode of two phonon vibrations. In anisotropic TMDCs such as $ReS_2$ (Figure 5d) [76, 77] and $ReSe_2$ [74, 78], more complex Raman response has been observed. Since there are 12 non-equivalent atoms in each $ReSe_2$ unit cell, the material has a total of 36 phonon modes, including 18 Raman active modes, of which three modes are ultra-low frequency acoustic modes and 15 are intra-layer vibration modes. More than ten peaks between 100 cm$^{-1}$ to 300 cm$^{-1}$ have been observed in the recent work of $ReSe_2$ Raman study [74, 78], which is consistent with the DFT calculations. All the modes have polarization dependent intensities (Figure 5d), which can be utilized to identify the crystal orientation of the flakes.

Inter-layer Raman modes is especially important for layered materials since not only can it be applied as a nondestructive method to identify layer numbers, but also carry useful information of interlayer interaction properties such as charge exchange, screenings and scatterings. For anisotropic materials, it can also be used to determine crystal orientations. Owning to the in-plane symmetry, the two shear modes in isotropic 2D materials are often degenerated to appear as a single peak in the Raman spectra. However, in anisotropic layered materials, the broken symmetry results in a split of the two shear modes, although it may not be well distinguished due to the experimental resolution, resulting in a peak broadening. In addition, due to the anisotropic in-plane crystal structure, the Raman peak intensity of inter-layer modes are sensitive to the polarization of incident light, hence it can be utilized to identify crystal orientation. Recently, the interlayer Raman modes of ReSe$_2$ [74] and BP [79] flakes has been explored both theoretically and experimentally (Figure 5a, 5b and 5c). Polarization dependence of the low-frequency modes has been detected and the linear chain model has been proven to be powerful in predicting mode positions.

**New Opportunities for Piezoelectric Devices**

Besides the interesting features in carrier transport, optical response and phonon activity, low symmetry 2D materials also offer unique characteristics in their piezoelectric properties.Piezoelectricity is the electric polarization charge that accumulates in solids in response to the applied mechanical stress/pressure [80-82]. Many materials exhibit the piezoelectric effect. For example, non-centrosymmetric wurtzite-structured semiconductors, such as ZnO, GaN, and InN, have been widely used in piezotronic and piezo-phototronic devices [83-85]. 2D materials, including transition metal dichalcogenides (TMDCs) and h-BN, have recently sparked new interests for piezoelectric applications [86-91] because of their high crystallinity and the ability to withstand enormous strain [92, 93]. Monolayer group III monochalcogenides may also exhibit the similar piezoelectric effect as TMDCs due to the broken inversion symmetry. However, the piezoelectric effects of these 2D structures are rather weak and the mechanical-electrical energy converting efficiency is limited. For example, the piezoelectric coefficient $d_{11}$ of monolayer MoS$_2$ is around 2.9x10$^{-10}$ C/m [90], and the corresponding mechanical-electrical energy conversion rate is limited to be about 5% [89]. Although there are enhanced piezoelectric materials, e.g. lead zirconium titanate Pb [Zr$_x$Ti$_{1-x}$]O$_3$ (PZT) [83, 94, 95], they are not 2D structures and their brittle nature causes application problems and severely limits the amount of strain that can be sustained [96].

Comparing to other 2D materials, group IV monochalcogenides may offer enhanced piezoelectric properties because of their low level of in-plane crystal symmetry and anisotropic characteristics (Figure 1, lower panel). Figure 6 compares the piezoelectric coefficients of group IV monochalcogenides to that of other 2D structures. First, these materials are non-centrosymmetric, giving rise to piezoelectric effects; second, the unique puckered structure ($C_{2v}$ symmetry) makes them very soft along the armchair direction, possibly further enhancing piezoelectric efficiency induced by strain.Employing the established first principles DFT calculations based on Berry phase, it is predicted that these monolayer monochalcogenides exhibit surprisingly strong piezoelectric effects [5]. As shown in Figure 6, the characteristic piezoelectric coefficients $d_{11}$ is about two orders of magnitude larger than those of other 2D materials, such as $MoS_2$ [86], GaS [93], bulk quartz and AlN [83]. Furthermore, because of the anisotropic puckered structures, the piezoelectric effect is highly anisotropic. Along the zigzag direction, because of the mirror symmetry, the piezoelectric effects is zero. A more recent work focusing on electronic structures of these materials also observes the similar giant piezoelectric effect [97]. The predicted giant piezoelectricity in those monolayer group IV monochalcogenides is from a direct *ab initio* calculation [5]. However, the deep physics has not been well understood. The naive idea is to attribute the enhanced piezoelectricity to the softer structures. However, according to the calculation, the elastic stiffness difference between these monolayer group IV chalcogenides with other 2D materials is only about one order of magnitude [98]. Thus this mechanical reason is not enough to explain the two orders of magnitude enhancement. Deeper reasons and new physics may be expected. For example, the polarization of these group IV monochalcogenides exhibit some nonlinear response to the external strain, as noticed in Ref. [5]. The further research about the memory effect under external electric field [99] and ferroelectricity of monolayer group IV monochalcogenides shows the anharmonic potential existing in these materials [100]. In particular, Monte Carlo simulations [100] reveal that the Curie temperature of these 2D ferroelectric materials are well above room temperature, making them useful for device applications. The anharmonic potential causes strong phonon-phonon interaction, inducing the short lifetime of phonon. Thus the anharmonic effect could induce the ferroelectric instability in ferroelectric material [100], which also result in large piezoelectric coefficients. Meanwhile, the anharmonic potential caused by the short lifetimes of phonons could result in very low thermal conductance, which is favorable for thermoelectric applications. The strong piezoelectric properties of low-symmetry 2D materials can find many applications not only in piezoelectric energy harvesting, but also in radio frequency (RF) signal processing as surface acoustic wave devices and devices with non-reciprocal characteristics for filtering, transmission and amplification applications. Comparing to incumbent approaches for these applications based on traditional piezoelectric materials and the emerging nitride material family, 2D materials offer unprecedented mechanical flexibility, decent mobility and the ease of integration with standard silicon CMOS processes. On the other hand, the research in the piezoelectric application of low symmetry 2D materials is still in its early stage and many challenges exist such as the scalable synthesis of these materials and the control of the layer number uniformity and the crystalline quality across a large area.

Beyond the fundamental research and the existing obstacles in material synthesis, the main challenge for the piezoelectric nanomaterials in realistic applications may also come from the limit of few-layer structures because the inversion symmetry will be restored in even-number-layer structures. In other words, the strain-induced polarization field will cancel each other by neighboring layers. For odd-number-layer structures, the useful piezoelectric effect is essentially from the residual layer. Thus the average power per thickness is reduced for multi-layer structures. How to overcome the limit of output power will be crucial for realistic device applications. There are ways to potentially solve this problem. For instance, chemical intercalations or van der Waals (vdW) heterojunctions may break the inversion symmetry of multilayer structures and they are helpful for keeping the piezoelectricity. The output power for applications such as energy harvesting is also decided by the internal resistance of piezoelectric materials. Unfortunately, we have very limited reliable knowledge about the free-carrier mobility in monolayer/few-layer group IV monochalcogenides to date. This may also be a fundamental challenge because their giant piezoelectric effect may induce significant acoustic phonon-electron interactions (scatterings) and reduce the free-carrier mobility. The surface and substrate scatterings will also impact the conductance significantly. More novel ideas are necessary for overcoming this problem in the future.

To date, most of the research interest has been focused on the in-plane piezoelectric effect; the in-plane polarization is induced by the in-plane strain. It is obvious that the off-plane piezoelectric effect, in which the vertical electrical field can be generated by the in-plane strain, may be useful for device applications. For example, in the widely used FET structures, instead of applying a gating voltage, the strain can generate an off-plane polarization field working as the gating field to tune the device performance. In fact, some 2D materials have been predicted to exhibit this off-plane piezoelectric effect. As shown in layered III-V semiconductors [92], their off-plane piezoelectric coefficients can be similar to those of in-plane ones, making them useful for piezoelectric transistors.

Finally, as mentioned in the earlier sections, the prospect for realistic piezoelectric applications relies on the success of developing robust techniques to protect the structural stability of these anisotropic 2D structures under ambient conditions. In addition, the oxidization process is rather easily associated with the unique lone pair electronic structure, defects and straining condition in BP [101-104]. In the future, it is of interest to check if these mechanisms are also effective in explaining the stability of the isoelectronic group IV monochalcogenides. Meanwhile, how those anisotropic structures evolve with temperature and if there is any possible phase transition related to their anisotropic, low symmetry crystal structure are also crucial in this field.

**Prospects for Thermoelectric Applications**

The low symmetry crystal structure that offers interesting piezoelectric properties in group IV monochalcogenides as discussed in the previous section may also benefit thermoelectric applications. Thermoelectric devices utilize the Seebeck effect to convert heat flow into electrical energy and are highly desirable for the development of solid state, passively powered electronic systems. The efficiency of thermoelectric materials is measured by its figure of merit, $ZT = S^2 \frac{\sigma}{\kappa} T$, where $S$ is the Seebeck coefficient, $\sigma$ is the electrical conductivity, $\kappa$ is the thermal conductivity, and $T$ is the temperature. The forefront of all-scale electronic and atomistic structural engineering techniques have served to enhance the $\sigma/\kappa$ ratios of third-generation thermoelectric materials, such as Cubic AgPb$_m$SbTe$_{2+m}$ [105], PbTe-SrTe [106, 107] and In$_4$Se$_{3-\delta}$ [108], achieving ZT values near 2 (around a 15% conversion efficiency) within a temperature range of 700 K to 900 K.

Meanwhile, significant efforts have been performed to look for high-efficiency thermoelectric materials with reduced dimensionalities. Promising figure of merit ZT have been achieved in single-component nanowires [109] and multi-component thin film superlattices or quantum dots [105, 110, 111]. In particular, layered structures, such as SnSe, exhibit impressive peak ZT of 2.6 [4] (Figure 7a). Since black phosphorus has a similar structure as SnSe, it immediately ignites interest in its thermoelectric properties. For example, large thermoelectric power factor has been predicted based on an initio DFT calculations based on electronic transport [112] and promising thermoelectric power has been observed in bulk black phosphorus [113].

Another approach for studying thermoelectrics is motivated by the anisotropic effect on thermoelectric efficiency. As shown in bulk SnSe, the anharmonic and anisotropic bonds [114] are expected to enhance the thermoelectric efficiency [115]. Considering the observed anisotropic electrical conductance in black phosphorous, it is obvious to check if the lattice thermal conductance also exhibits anisotropy. A first-principles calculation based on the Bardeen and Shockley model shows that the lattice thermal conductance is anisotropic [98]. More sophisticated simulations based on the Boltzmann equations derived a similar conclusion [116, 117]. This anisotropic lattice thermal conductance of few-layer black phosphorus and their nanoribbon structures [118] has been confirmed by recent experiments. The electron-photon and phonon-phonon interactions are proposed to be anisotropic as well [98]. More phosphorus allotropes and the strain effects have been studied on their thermal conductance [117, 119].

Interestingly, the preferred thermal conductance direction of monolayer black phosphorous is orthogonal to the electrical conductance [98] (Figure 7b), enhancing the thermoelectric figure of merit because the thermoelectric conversion efficiency is proportional to the ratio of a device's electrical conductance to its thermal conductance. As a result, the theoretical study estimates that the ZT of monolayer black phosphorus can reach 1 under a moderate doping condition at room temperature, making it promising for thermoelectric applications (Figure 7b). Beyond intrinsic black phosphorus, broader research on layered phosphorus allotropes and strain effects have also been carried out. For example, ZT of black phosphorus can be further increased to 5.4 by substituting the P atom with the Sb atom, giving a nominal formula of $P_{0.75}Sb_{0.25}$ [120]; a 10% strain on monolayer black phosphorus can exhibit a very large power factor exceeding 10 mW/(m.K$^2$) by DFT calculations [121]. The relevant spin-Seebeck effect is also proposed for nanoribbon structures of black phosphorus [122].

The 2D structure and low-symmetry induced anisotropy will impact the thermoelectric performance significantly. Hence, it is of fundamental interests to study the thermoelectric behavior of the few-layer structures. For example, the anharmonic effect of lattice vibrations in bulk SnSe is proposed to be crucial for enhancing the ZT. In the few-layer and monolayer of these materials, how those anharmonic effects evolve with the quantum confinement and if this variation will enhance or reduce the thermoelectric performance are interesting problems awaiting further study. Apart from the Umklapp process, which describes the phonon-phonon interaction and will affect the thermal conductance, the phonon life time is also determined by the phonon and boundary scattering. In few layers and monolayer materials, the interaction between phonon and boundary might be enhanced and the phonon lifetime would be substantially reduced, causing potentially low thermal conductance. Hence, 2D materials might be good candidates for high performance thermoelectric devices.

Another potentially interesting topic will be how the vibrational modes impact the thermal conductance and thermoelectric efficiency in 2D structures. For example, as shown in previous studies of 2D structures on substrates [123], the off-plane lattice vibrational modes will be quenched in multilayer structures by off-plane interlayer interactions and change the thermal conductance significantly. Therefore, it is of particular interests to study the effects of those vibration modes on electron-phonon coupling and thermoelectric performance.

Quantum confinement effects on the band gap may be another factor to impact the thermoelectric efficiency. When the thickness is reduced to a few nm, the quantum confinement effect will significantly enlarge the bandgap. According to the calculation of the thermoelectric figure of merit, this bandgap enlargement in few-layer structures will change the electrical transport by varying the effective mass and thermal excited free-carrier number. This will ultimately vary thermoelectric performance as well. This factor may have to be considered when studying the thermoelectric performance of monolayer/few-layer group IV monochalcogenides.

The unique vdW structure may also be useful for enhancing the thermoelectric performance. The enhancement of thermoelectric figure of merit is from the ratio between the electronic and thermal conductances. Unlike the bulk materials, in which all atoms are connected by strong covalent bonds, layered structures are formed by the weak interlayer vdW interactions. It is obvious that the thermal and electronic transport across these vdW interfaces will be different. For example, the electronic free carriers may have to transport across the vdW interface through the tunneling effect while the phonon transport will go through the vdW interfaces. Thus the ratio between these two types of transport may be different from usually covalent interfaces. Following this idea, applying perpendicular pressure to tune interlayer distance and the vdW potential barriers may dramatically change the off-plane transport properties, giving hope to tuning the thermoelectric properties.

The thermoelectric effects in nanoribbon structures of these anisotropic materials will also be interesting. For example, as widely discussed in graphene nanoribbons, the edge structure may influence the electrical and thermal conductance in different ways, and change their ratio. As noticed in many other 1D structures, these modifications may enhance the thermoelectric efficiency [124, 125]. In particular, for these anisotropic 2D structures (black phosphorus and group IV monochalcogenides), the orientation of these nanoribbons exhibit different effective mass and bandgaps. This will impact their transport and thermoelectric properties.

These unique thermoelectric effect in anisotropic monochalcogenides may bring exciting applications if combined with their excited state properties. These materials have been proposed to be useful for valleytronics and optical excitations [126]. The thermal temperature gradient may bring a polarization field to the 2D structures. It is interesting to drive the Hall Effect by thermal power instead of applied external electric field, resulting in a thermoelectric Hall Effect.

Finally, despite numerous theoretical calculations on thermoelectric properties of 2D anisotropic materials, the reliability of those calculations is still questionable. This uncertainty comes from the complicated nature of thermoelectric effects. For example, in most studies, the crucial scattering time (phonon/electron), which decides the conductance, is from either fitting experimental bulk values or assuming constant scattering time approximation. However, in realistic few-layer structures, the scattering time must be different from bulk values. More importantly, the scattering time shall be a dynamic factor, instead of a constant in the whole energy spectrum. Additionally, the surface and substrate scattering is very important for ultra-thin structures. Unfortunately, these scattering mechanisms have not been well included in the current simulations. More sophisticated and fundamental advances in simulation approaches for predicting transport properties are important for quantitatively guiding experimental efforts and this may go beyond the scope of studying anisotropic, low-symmetry 2D materials.

**Conclusion**

In this review, we discussed the unique properties of low-symmetry 2D materials arising from their low-symmetry crystal lattice, the methods for synthesizing these materials, and their applications in innovative devices for electronics and photonics. We also provided an outlook of the promising future directions in low-symmetry 2D materials research and offered our perspective on the emerging device applications these materials may enable. The unique properties arising from their reduced symmetry can benefit many application fields in electronics and photonics, as well as piezoelectronics and thermoelectrics. As discussed in the previous sections, the anisotropic physical nature of these materials and the unique properties arising from their low symmetry crystal structure are promising for many potential applications from high efficiency thermoelectric devices for energy harvesting to non-reciprocal RF and acoustic devices for wireless communication systems, as well as for bio-inspired electronics and optical computing. There will certainly be many challenges leading to the eventual applications of these materials, from scalable synthesis and fundamental property study to device innovation and system implementation. On the other hand, there exist a vast array of exciting opportunities that merit this emerging family of materials extensive scientific investigations and technological development in the years to come.

**Figures and Captions**

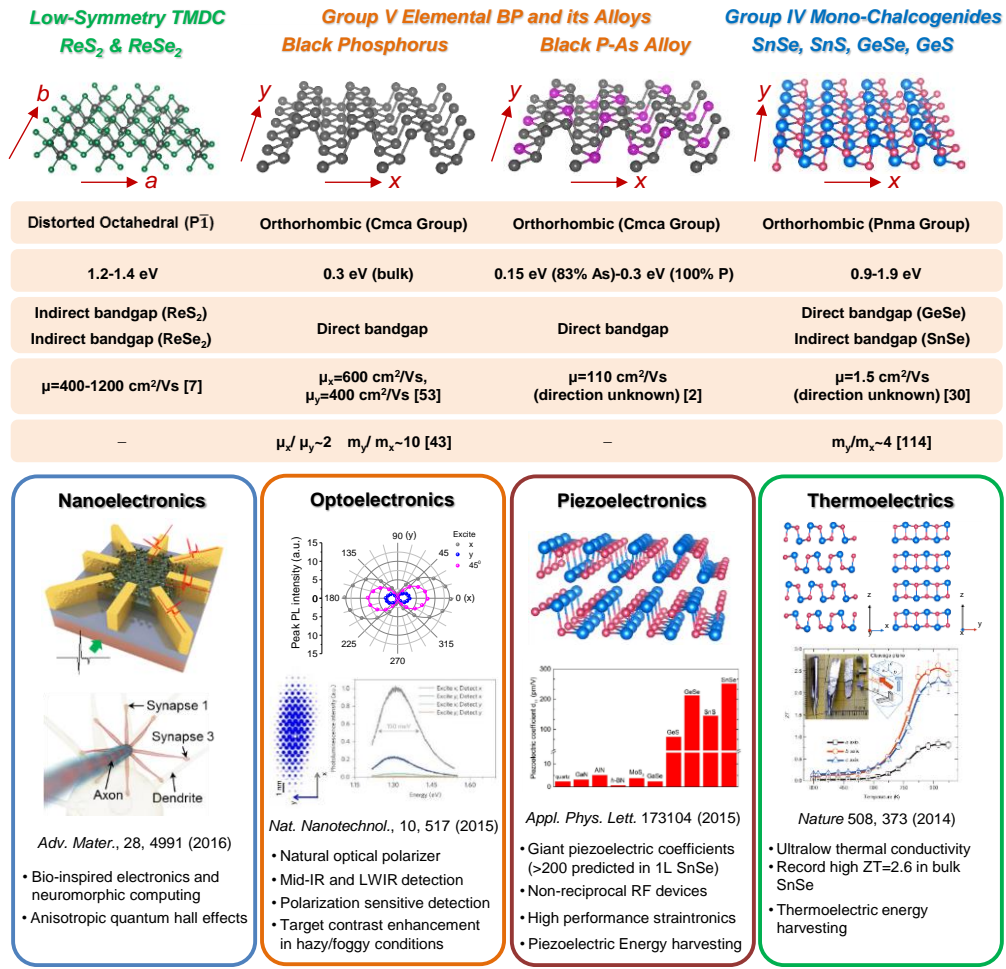

**Figure 1. The crystal structures, basic properties and device applications of low symmetry 2D materials.** Upper panel: the crystal structures of the various types of low symmetry 2D materials; the table lists their space group, bandgap, bandgap type, mobility, and in-plane mobility ratio information from the top to the bottom rows, respectively. Lower panel: the potential applications in electronic, photonic, piezoelectric and thermoelectric devices that can be enabled by low symmetry 2D materials, which are discussed in detail in later sections of this article. Figures in the bottom row are modified with permission from [21] Copyright 2016 Wiley-VCH, [69] Copyright 2015 Nature Publishing Group, [5] Copyright 2015 AIP Publishing LLC, [4] Copyright 2014 Nature Publishing Group.

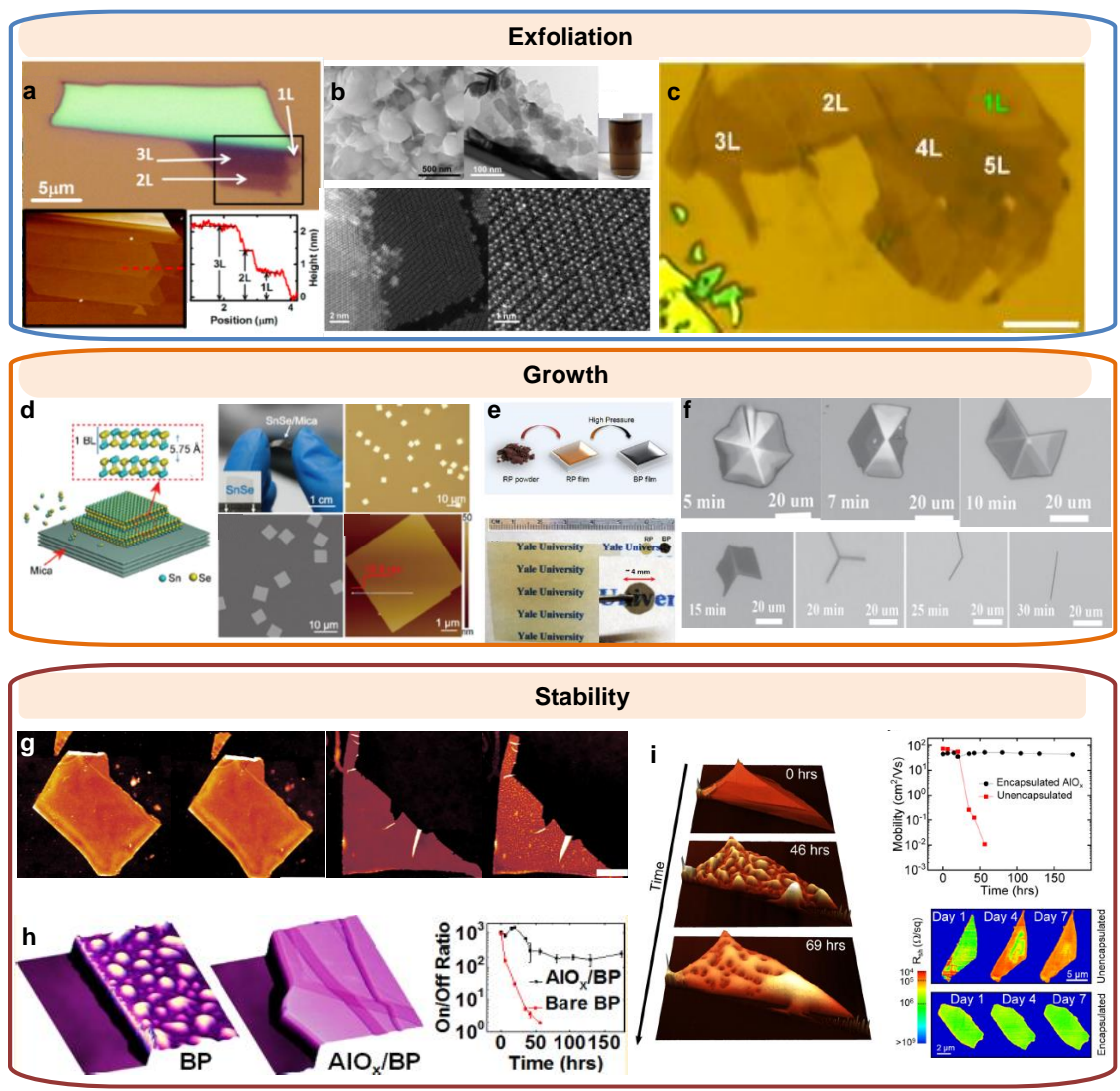

**Figure 2. Low symmetry 2D materials exfoliation, synthesis and stability.** (a) Mechanically exfoliated single and few layer ReS$_2$. (b) Chemically exfoliated ReS$_2$ nanosheets from powder. (c) Plasma thinning of exfoliated BP. (d) Vapor transport grown SnSe nanosheets. (e) High pressure synthesis of BP thin films on flexible substrates. (f) CVD of ReS$_2$ layers and nanoribbons as a function of growth time. (g) AFM images of chemically passivated BP over 10 days (left) and no passivation under ambient conditions (right). (h) AlO$_x$ encapsulation can protect BP from degradation and maintain the on/off ratio and (i) PMMA dielectric capping to preserve high mobility in BP FETs. Modified with permission from (a) [76] Copyright 2015 American Chemical Society, (b) [40] Copyright 2014 The Royal Society of Chemistry,

(c) [127] Copyright 2014 Tsinghua University Press and Springer-Verlag Berlin Heidelberg, (d) [30] Copyright 2015 Tsinghua University Press and Springer-Verlag Berlin Heidelberg, (e) [9] Copyright 2015 IOP PUBLISHING, (f) [28], Copyright 2015 Wiley-VCH Verlag, (g) [23] Copyright 2016 Nature Publishing Group, (h) [17] Copyright 2014 American Chemical Society (i) [128] 2015 American Chemical Society.

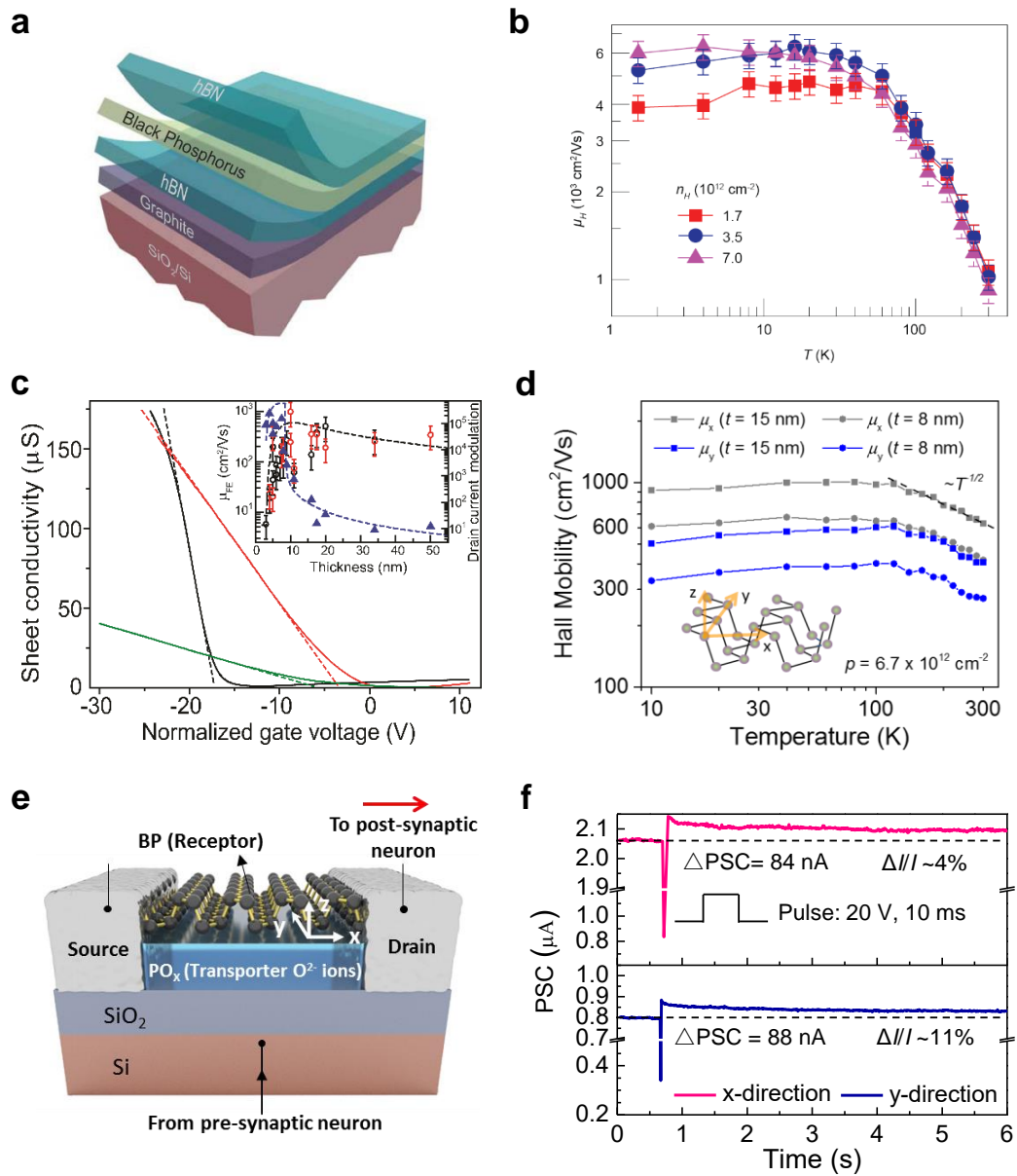

**Figure 3. Anisotropic transport properties and electronic applications of BP.** (a) Schematic of BP device with the BP channel sandwiched between two h-BN thin

films and (b) the measured hall mobility of a BP sample protected in this sandwich structure. (c) Sheet conductivity measured as a function of the gate voltage for BP devices with different channel thicknesses: 10 nm (black solid line), 8 nm (red solid line) and 5 nm (green solid line), with field-effect mobility values of 984, 197 and 55 cm$^2$ V$^{-1}$ s$^{-1}$, respectively. Field-effect mobilities were extracted from the line fit of the linear region of the conductivity (dashed lines). Inset: drain current modulation (filled blue triangles) and carrier mobility (open circles) extracted for BP FETs with different channel thicknesses. (d) Hall mobility of BP thin films in both x- and y-directions. (e) The schematic of a BP synaptic device. (f) Anisotropic synaptic responses of BP synapses fabricated along the x- and y-directions of the BP crystal. Modified with permission from (a and b) [47] Copyright 2016 Nature Publishing Group, (c) [51] Copyright 2014 Nature Publishing Group, (d) [53] Copyright 2014 Nature Publishing Group, (e and f) [21] Copyright 2016 Wiley-VCH.

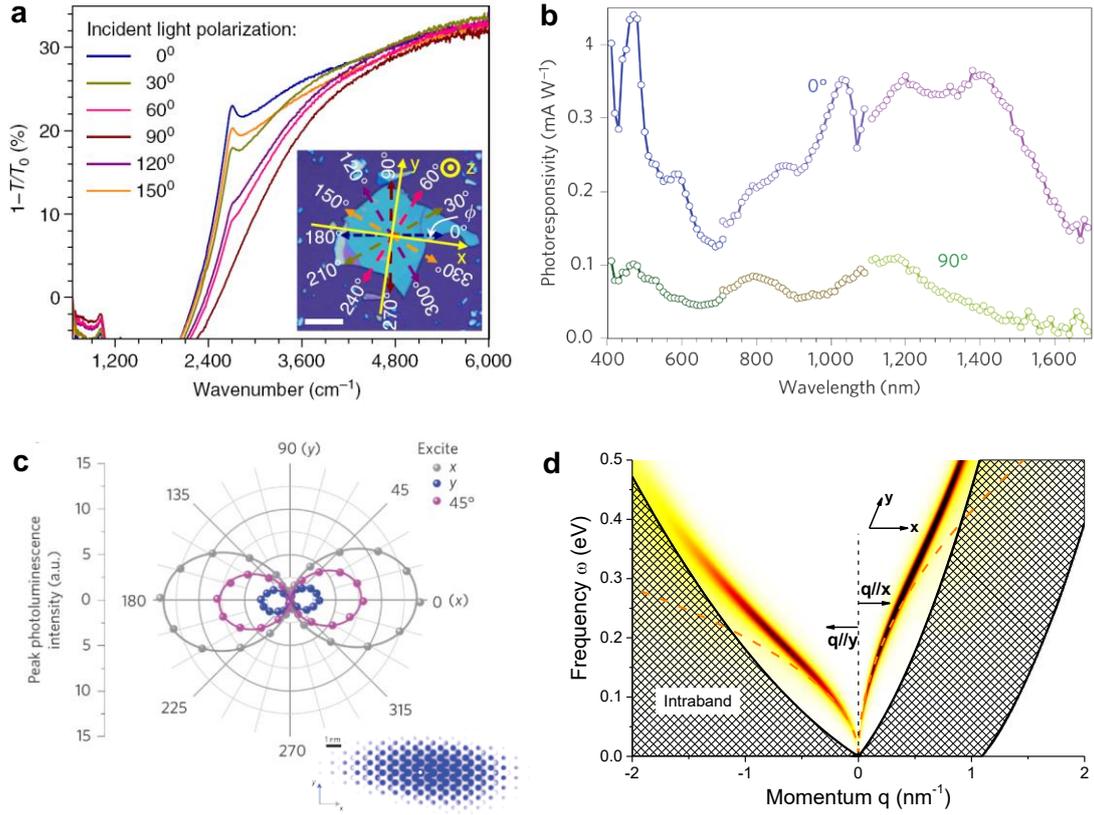

**Figure 4. Anisotropic optical properties and photonic applications of BP.** (a) Polarization-resolved relative extinction spectra of thin film BP in the infrared wavelength range. (b) The wavelength spectra of the photo-responsivity for a BP photodetector in the infrared range for incident light polarized along x- and y-directions of the BP crystal. (c) Peak photoluminescence intensity of monolayer BP vs. the polarization detection angle for incident excitation laser linearly polarized along different crystal directions. The inset shows the wave function of the carriers in monolayer BP. (d) Theoretical prediction of the anisotropic plasmon dispersion along the x- and y-directions of monolayer BP. The shaded areas indicate the Landau damping regions. Modified with permission from (a) [53] Copyright 2014 Nature Publishing Group, (b) [60] Copyright 2015 Nature Publishing Group, (c) [69] Copyright 2015 Nature Publishing Group, (d) [56] Copyright 2014 American Physical Society.

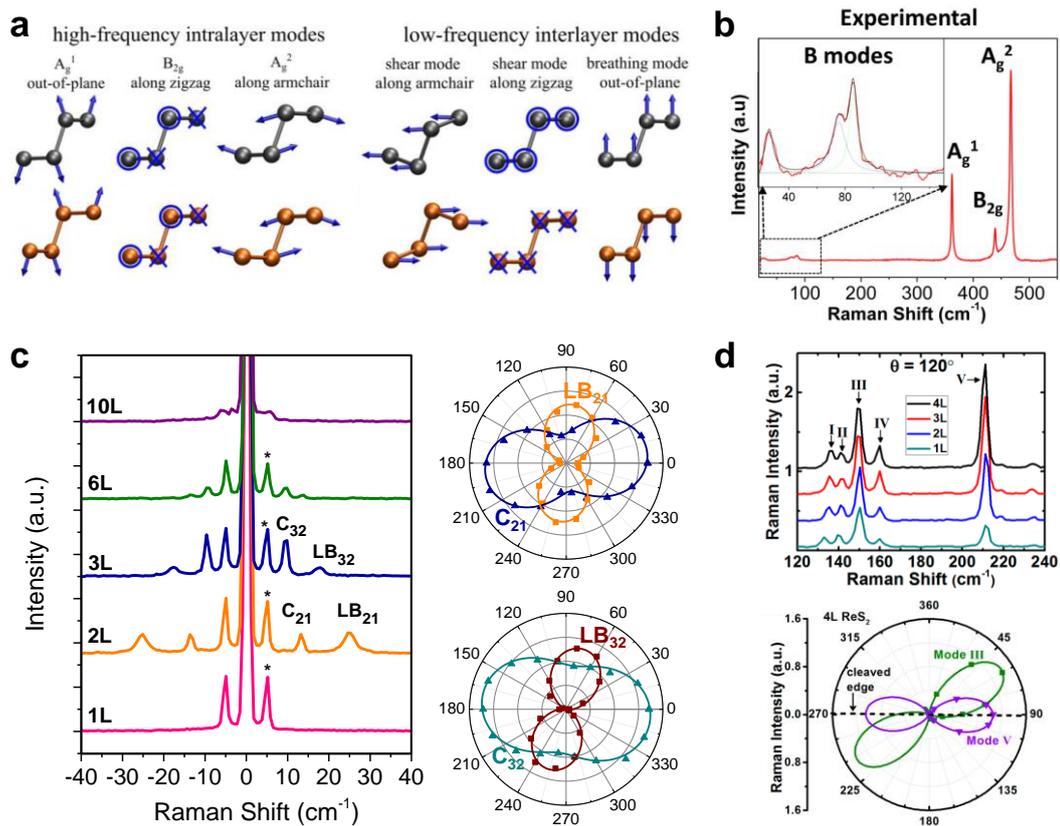

**Figure 5. Anisotropic Raman scattering in low-symmetry 2D materials.** (a) The schematic of the intra- and inter-layer Raman vibration modes in BP. (b) The typical Raman spectra of BP. (c) Anisotropic low-frequency interlayer vibration modes in atomically-thin ReSe$_2$. (d) The typical Raman spectra of single- and few-layer ReS$_2$. The dependence of the III and V Raman mode intensities in ReS$_2$ on the linear polarization of the excitation light. Modified with permission from (a,b) [79] Copyright 2015 American Chemical Society, (c) [74] Copyright 2015 Tsinghua University Press and Springer-Verlag Berlin Heidelberg, (d) [76] Copyright 2015 American Chemical Society.

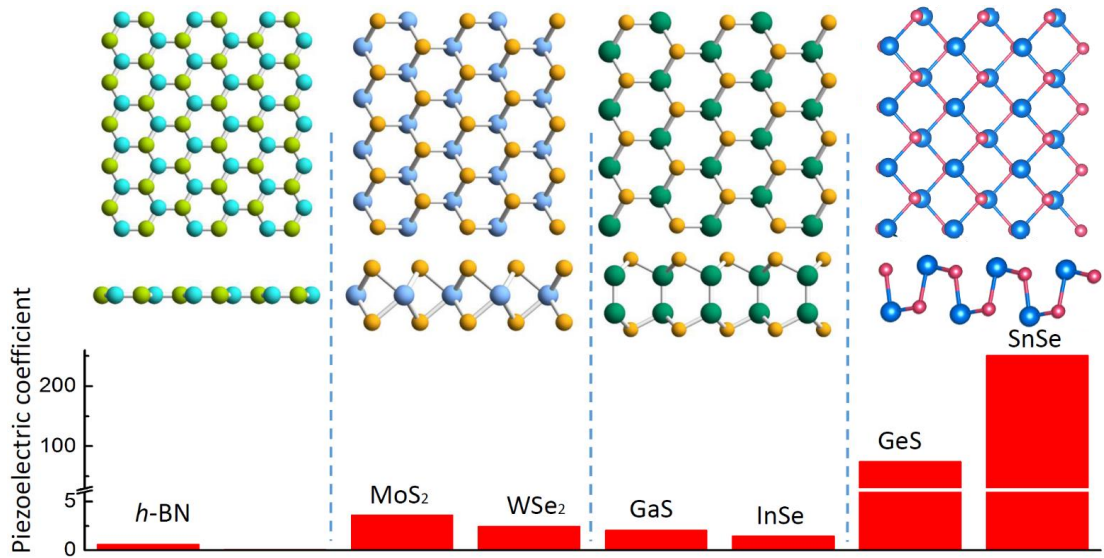

**Figure 6. The characteristic piezoelectric coefficients ($d_{11}$(pm/V)) of typical layered 2D materials.** The upper panels are the top and side views of the ball-stick models. Modified with permission from [5] Copyright 2015 AIP Publishing LLC.

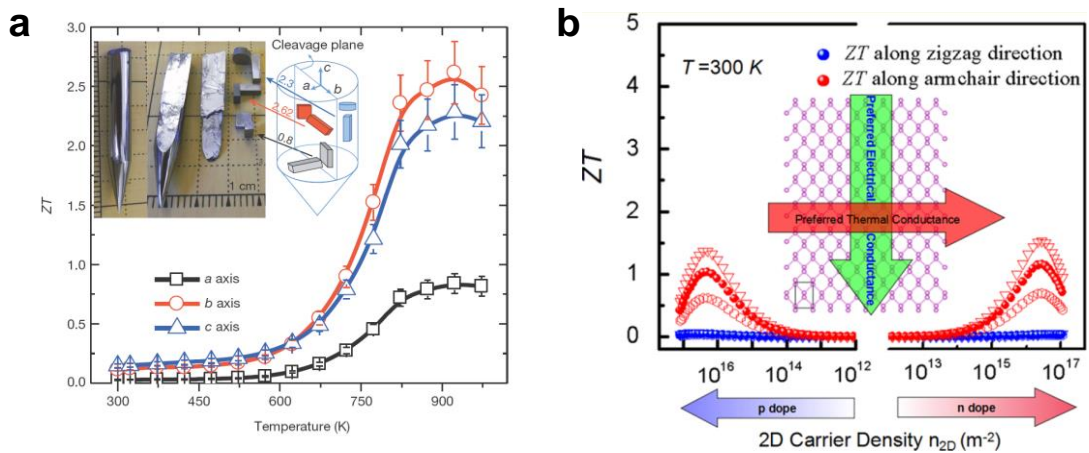

**Figure 7. The thermoelectric properties of low symmetry materials.** (a) The experimentally measured ZT factor of bulk SnSe along x-, y- and z-directions under different temperature. (b) The ZT factor along x- and y-directions in monolayer BP at room temperature. Modified with permission from (a) [4] Nature Publishing Group, (b) [98] American Chemical Society.

# References

[1] J. Kim, S.S. Baik, S.H. Ryu, Y. Sohn, S. Park, B.-G. Park, J. Denlinger, Y. Yi, H.J. Choi, K.S. Kim, Science, 349 (2015) 723-726.

[2] B. Liu, M. Köpf, A.N. Abbas, X. Wang, Q. Guo, Y. Jia, F. Xia, R. Weihrich, F. Bachhuber, F. Pielnhofer, Advanced Materials, 27 (2015) 4423-4429.

[3] A. Agarwal, M. Vashi, D. Lakshminarayana, N. Batra, Journal of Materials Science: Materials in Electronics, 11 (2000) 67-71.

[4] L.-D. Zhao, S.-H. Lo, Y. Zhang, H. Sun, G. Tan, C. Uher, C. Wolverton, V.P. Dravid, M.G. Kanatzidis, Nature, 508 (2014) 373-377.

[5] R. Fei, W. Li, J. Li, L. Yang, Applied Physics Letters, 107 (2015) 173104.

[6] G. Shi, E. Kioupakis, Nano letters, 15 (2015) 6926-6931.

[7] E. Liu, Y. Fu, Y. Wang, Y. Feng, H. Liu, X. Wan, W. Zhou, B. Wang, L. Shao, C.-H. Ho, Nature communications, 6 (2015).

[8] P. Bridgman, Journal of the American Chemical Society, 36 (1914) 1344-1363.

[9] X. Li, B. Deng, X. Wang, S. Chen, M. Vaisman, S.-i. Karato, G. Pan, M.L. Lee, J. Cha, H. Wang, 2D Materials, 2 (2015) 031002.

[10] T. Nilges, M. Kersting, T. Pfeifer, Journal of Solid State Chemistry, 181 (2008) 1707-1711.

[11] M. Zhao, Y. Wang, H. Qian, X. Niu, W. Wang, L. Guan, J. Sha, Crystal Growth & Design, (2016).

[12] M. Köpf, N. Eckstein, D. Pfister, C. Grotz, I. Krüger, M. Greiwe, T. Hansen, H. Kohlmann, T. Nilges, Journal of Crystal Growth, 405 (2014) 6-10.

[13] J. Jia, S.K. Jang, S. Lai, J. Xu, Y.J. Choi, J.-H. Park, S. Lee, ACS nano, 9 (2015) 8729-8736.

[14] K.-Y.C. James Nelson Baillargeon, Alfred Yi Cho, Sung-Nee George Chu, Wen-Yen Hwang, US 6110438 A (2000).

[15] O. Osters, T. Nilges, F. Bachhuber, F. Pielnhofer, R. Weihrich, M. Schöneich, P. Schmidt, Angewandte Chemie International Edition, 51 (2012) 2994-2997.

[16] J.-S. Kim, Y. Liu, W. Zhu, S. Kim, D. Wu, L. Tao, A. Dodabalapur, K. Lai, D. Akinwande, Scientific reports, 5 (2015).

[17] J.D. Wood, S.A. Wells, D. Jariwala, K.-S. Chen, E. Cho, V.K. Sangwan, X. Liu, L.J. Lauhon, T.J. Marks, M.C. Hersam, Nano letters, 14 (2014) 6964-6970.

[18] J.O. Island, G.A. Steele, H.S. van der Zant, A. Castellanos-Gomez, 2D Materials, 2 (2015) 011002.

[19] A. Favron, E. Gaufrès, F. Fossard, A.-L. Phaneuf-L'Heureux, N.Y. Tang, P.L. Lévesque, A. Loiseau, R. Leonelli, S. Francoeur, R. Martel, Nature materials, 14 (2015) 826-832.

[20] A. Castellanos-Gomez, L. Vicarelli, E. Prada, J.O. Island, K. Narasimha-Acharya, S.I. Blanter, D.J. Groenendijk, M. Buscema, G.A. Steele, J. Alvarez, 2D Materials, 1 (2014) 025001.

[21] H. Tian, Q. Guo, Y. Xie, H. Zhao, C. Li, J.J. Cha, F. Xia, H. Wang, Advanced Materials, (2016).

[22] R.A. Doganov, E.C. O'Farrell, S.P. Koenig, Y. Yeo, A. Ziletti, A. Carvalho, D.K. Campbell, D.F. Coker, K. Watanabe, T. Taniguchi, Nature communications, 6 (2015).

[23] C.R. Ryder, J.D. Wood, S.A. Wells, Y. Yang, D. Jariwala, T.J. Marks, G.C. Schatz, M.C. Hersam, Nature chemistry, (2016).

[24] C. Winkler, J. Prakt. Chem., 142 (1886) 215.

[25] W.Z. Hofmann, Kristallogr. Kristallgeom., Kristallphys., Kristallchem., 92A (1935) 161.

[26] N. Koteeswara Reddy, M. Devika, E. Gopal, Critical Reviews in Solid State and Materials

Sciences, 40 (2015) 359-398.

[27] A. Okazaki, I. Ueda, Journal of the Physical Society of Japan, 11 (1956) 470-470.

[28] A. Okazaki, Journal of the Physical Society of Japan, 13 (1958) 1151-1155.

[29] S. Badrinarayanan, A. Mandale, V. Gunjikar, A. Sinha, Journal of materials science, 21 (1986) 3333-3338.

[30] S. Zhao, H. Wang, Y. Zhou, L. Liao, Y. Jiang, X. Yang, G. Chen, M. Lin, Y. Wang, H. Peng, Nano Research, 8 (2015) 288-295.

[31] L. Li, Z. Chen, Y. Hu, X. Wang, T. Zhang, W. Chen, Q. Wang, Journal of the American Chemical Society, 135 (2013) 1213-1216.

[32] C. Li, L. Huang, G.P. Snigdha, Y. Yu, L. Cao, ACS nano, 6 (2012) 8868-8877.

[33] J. Xia, X.-Z. Li, X. Huang, N. Mao, D.-D. Zhu, L. Wang, H. Xu, X.-M. Meng, Nanoscale, 8 (2016) 2063-2070.

[34] J.R. Brent, D.J. Lewis, T. Lorenz, E.A. Lewis, N. Savjani, S.J. Haigh, G. Seifert, B. Derby, P. O'Brien, Journal of the American Chemical Society, 137 (2015) 12689-12696.

[35] K. Ramasamy, V.L. Kuznetsov, K. Gopal, M.A. Malik, J. Raftery, P.P. Edwards, P. O'Brien, Chemistry of Materials, 25 (2013) 266-276.

[36] H.V.A. Briscoe, Robinson, P.L., Stoddart, E.M., J. Chem. Soc., (1931) 1439.

[37] J. Wildervanck, F. Jellinek, Journal of the Less Common Metals, 24 (1971) 73-81.

[38] X. He, F. Liu, P. Hu, W. Fu, X. Wang, Q. Zeng, W. Zhao, Z. Liu, Small, 11 (2015) 5423-5429.

[39] K. Keyshar, Y. Gong, G. Ye, G. Brunetto, W. Zhou, D.P. Cole, K. Hackenberg, Y. He, L. Machado, M. Kabbani, Advanced Materials, 27 (2015) 4640-4648.

[40] T. Fujita, Y. Ito, Y. Tan, H. Yamaguchi, D. Hojo, A. Hirata, D. Voiry, M. Chhowalla, M. Chen, Nanoscale, 6 (2014) 12458-12462.

[41] B. Kang, Y. Kim, C. Lee, APS Meeting Abstracts2016.

[42] Y. Akahama, S. Endo, S.-i. Narita, Journal of the Physical Society of Japan, 52 (1983) 2148-2155.

[43] J. Qiao, X. Kong, Z.-X. Hu, F. Yang, W. Ji, Nature communications, 5 (2014).

[44] L. Li, G.J. Ye, V. Tran, R. Fei, G. Chen, H. Wang, J. Wang, K. Watanabe, T. Taniguchi, L. Yang, Nature nanotechnology, 10 (2015) 608-613.

[45] X. Chen, Y. Wu, Z. Wu, Y. Han, S. Xu, L. Wang, W. Ye, T. Han, Y. He, Y. Cai, Nature communications, 6 (2015).

[46] N. Gillgren, D. Wickramaratne, Y. Shi, T. Espiritu, J. Yang, J. Hu, J. Wei, X. Liu, Z. Mao, K. Watanabe, 2D Materials, 2 (2014) 011001.

[47] L. Li, F. Yang, G.J. Ye, Z. Zhang, Z. Zhu, W. Lou, X. Zhou, L. Li, K. Watanabe, T. Taniguchi, Nature nanotechnology, (2016).

[48] H. Wang, D. Nezich, J. Kong, T. Palacios, IEEE Electron Device Letters, 30 (2009) 547-549.

[49] W. Zhu, M.N. Yogeesh, S. Yang, S.H. Aldave, J.-S. Kim, S. Sonde, L. Tao, N. Lu, D. Akinwande, Nano letters, 15 (2015) 1883-1890.

[50] H. Wang, A. Hsu, J. Wu, J. Kong, T. Palacios, (2010).

[51] L. Li, Y. Yu, G.J. Ye, Q. Ge, X. Ou, H. Wu, D. Feng, X.H. Chen, Y. Zhang, Nature nanotechnology, 9 (2014) 372-377.

[52] H. Liu, A.T. Neal, Z. Zhu, Z. Luo, X. Xu, D. Tománek, P.D. Ye, ACS nano, 8 (2014) 4033-4041.

[53] F. Xia, H. Wang, Y. Jia, Nature communications, 5 (2014).

[54] S.P. Koenig, R.A. Doganov, H. Schmidt, A.C. Neto, B. Oezyilmaz, Applied Physics Letters, 104 (2014) 103106.

[55] V.A. Fonoberov, A.A. Balandin, Nano letters, 6 (2006) 2442-2446.
[56] T. Low, R. Roldán, H. Wang, F. Xia, P. Avouris, L.M. Moreno, F. Guinea, Physical review letters, 113 (2014) 106802.
[57] H. Wang, X. Wang, F. Xia, L. Wang, H. Jiang, Q. Xia, M.L. Chin, M. Dubey, S.-j. Han, Nano letters, 14 (2014) 6424-6429.
[58] M. Buscema, D.J. Groenendijk, S.I. Blanter, G.A. Steele, H.S. van der Zant, A. Castellanos-Gomez, Nano letters, 14 (2014) 3347-3352.
[59] M. Buscema, D.J. Groenendijk, G.A. Steele, H.S. van der Zant, A. Castellanos-Gomez, Nature communications, 5 (2014).
[60] H. Yuan, X. Liu, F. Afshinmanesh, W. Li, G. Xu, J. Sun, B. Lian, A.G. Curto, G. Ye, Y. Hikita, Nature nanotechnology, 10 (2015) 707-713.
[61] M. Engel, M. Steiner, P. Avouris, Nano letters, 14 (2014) 6414-6417.
[62] N. Youngblood, C. Chen, S.J. Koester, M. Li, Nature Photonics, (2015).
[63] Q. Guo, A. Pospischil, M. Bhuiyan, H. Jiang, H. Tian, D. Farmer, B. Deng, C. Li, S.-J. Han, H. Wang, arXiv preprint arXiv:1603.07346, (2016).
[64] T.A.F. Germer, M. J., Proc. SPIE 518 (2003) 264.
[65] D. Miyazaki, R.T. Tan, K. Hara, K. Ikeuchi, Computer Vision, 2003. Proceedings. Ninth IEEE International Conference on, IEEE2003, pp. 982-987.
[66] S.A. Empedocles, R. Neuhauser, M.G. Bawendi, Nature, 399 (1999) 126-130.
[67] J.S. Tyo, M. Rowe, E. Pugh, N. Engheta, Applied optics, 35 (1996) 1855-1870.
[68] V. Tran, R. Soklaski, Y. Liang, L. Yang, Physical Review B, 89 (2014) 235319.
[69] X. Wang, A.M. Jones, K.L. Seyler, V. Tran, Y. Jia, H. Zhao, H. Wang, L. Yang, X. Xu, F. Xia, Nature nanotechnology, 10 (2015) 517-521.
[70] V. Tran, R. Fei, L. Yang, 2D Materials, 2 (2015) 044014.
[71] J. Yang, R. Xu, J. Pei, Y.W. Myint, F. Wang, Z. Wang, S. Zhang, Z. Yu, Y. Lu, Light: Science & Applications, 4 (2015) e312.
[72] L. Li, J. Kim, C. Jin, G. Ye, D.Y. Qiu, F.H. da Jornada, Z. Shi, L. Chen, Z. Zhang, F. Yang, arXiv preprint arXiv:1601.03103, (2016).
[73] F. Xia, Nature nanotechnology, 9 (2014) 575-576.
[74] H. Zhao, J. Wu, H. Zhong, Q. Guo, X. Wang, F. Xia, L. Yang, P. Tan, H. Wang, Nano Research, 8 (2015) 3651-3661.
[75] H. Li, Q. Zhang, C.C.R. Yap, B.K. Tay, T.H.T. Edwin, A. Olivier, D. Baillargeat, Advanced Functional Materials, 22 (2012) 1385-1390.
[76] D.A. Chenet, O.B. Aslan, P.Y. Huang, C. Fan, A.M. van der Zande, T.F. Heinz, J.C. Hone, Nano letters, 15 (2015) 5667-5672.
[77] S. Tongay, H. Sahin, C. Ko, A. Luce, W. Fan, K. Liu, J. Zhou, Y.-S. Huang, C.-H. Ho, J. Yan, Nature communications, 5 (2014).
[78] D. Wolverson, S. Crampin, A.S. Kazemi, A. Ilie, S.J. Bending, ACS nano, 8 (2014) 11154-11164.
[79] X. Ling, L. Liang, S. Huang, A.A. Puretzky, D.B. Geohegan, B.G. Sumpter, J. Kong, V. Meunier, M.S. Dresselhaus, Nano letters, 15 (2015) 4080-4088.
[80] A. Erturk, D.J. Inman, Piezoelectric Energy Harvesting, (2011) 1-18.
[81] K. Cook-Chennault, N. Thambi, A. Sastry, Smart Materials and Structures, 17 (2008) 043001.
[82] W. Heywang, K. Lubitz, W. Wersing, Piezoelectricity: evolution and future of a technology, Springer Science & Business Media2008.

[83] F. Bernardini, V. Fiorentini, D. Vanderbilt, Physical Review B, 56 (1997) R10024.
[84] I. Guy, S. Muensit, E. Goldys, Applied Physics Letters, 75 (1999) 4133-4135.
[85] Z.L. Wang, Advanced Materials, 24 (2012) 4632-4646.
[86] K.-A.N. Duerloo, M.T. Ong, E.J. Reed, The Journal of Physical Chemistry Letters, 3 (2012) 2871-2876.
[87] M.T. Ong, E.J. Reed, ACS nano, 6 (2012) 1387-1394.
[88] J. Qi, X. Qian, L. Qi, J. Feng, D. Shi, J. Li, Nano letters, 12 (2012) 1224-1228.
[89] W. Wu, L. Wang, Y. Li, F. Zhang, L. Lin, S. Niu, D. Chenet, X. Zhang, Y. Hao, T.F. Heinz, Nature, 514 (2014) 470-474.
[90] H. Zhu, Y. Wang, J. Xiao, M. Liu, S. Xiong, Z.J. Wong, Z. Ye, Y. Ye, X. Yin, X. Zhang, Nature nanotechnology, 10 (2015) 151-155.
[91] J. Qi, Y.-W. Lan, A.Z. Stieg, J.-H. Chen, Y.-L. Zhong, L.-J. Li, C.-D. Chen, Y. Zhang, K.L. Wang, Nature communications, 6 (2015).
[92] M.N. Blonsky, H.L. Zhuang, A.K. Singh, R.G. Hennig, ACS nano, 9 (2015) 9885-9891.
[93] W. Li, J. Li, Nano Research, 8 (2015) 3796-3802.
[94] R. Resta, D. Vanderbilt, Theory of polarization: a modern approach,  Physics of Ferroelectrics, Springer2007, pp. 31-68.
[95] T.D. Nguyen, N. Deshmukh, J.M. Nagarah, T. Kramer, P.K. Purohit, M.J. Berry, M.C. McAlpine, Nature nanotechnology, 7 (2012) 587-593.
[96] S.R. Anton, H.A. Sodano, Smart materials and Structures, 16 (2007) R1.
[97] L.C. Gomes, A. Carvalho, A.C. Neto, Physical Review B, 92 (2015) 214103.
[98] R. Fei, A. Faghaninia, R. Soklaski, J.-A. Yan, C. Lo, L. Yang, Nano letters, 14 (2014) 6393-6399.
[99] P.Z. Hanakata, A. Carvalho, D.K. Campbell, H.S. Park, arXiv preprint arXiv:1603.00450, (2016).
[100] R. Fei, W. Kang, L. Yang, arXiv preprint arXiv:1604.00724, (2016).
[101] A. Ziletti, A. Carvalho, P. Trevisanutto, D. Campbell, D. Coker, A.C. Neto, Physical Review B, 91 (2015) 085407.
[102] A. Ziletti, A. Carvalho, D.K. Campbell, D.F. Coker, A.C. Neto, Physical review letters, 114 (2015) 046801.
[103] Y. Liu, F. Xu, Z. Zhang, E.S. Penev, B.I. Yakobson, Nano letters, 14 (2014) 6782-6786.
[104] K.L. Utt, P. Rivero, M. Mehboudi, E.O. Harriss, M.F. Borunda, A.A. Pacheco SanJuan, S. Barraza-Lopez, ACS Central Science, 1 (2015) 320-327.
[105] K.F. Hsu, S. Loo, F. Guo, W. Chen, J.S. Dyck, C. Uher, T. Hogan, E. Polychroniadis, M.G. Kanatzidis, Science, 303 (2004) 818-821.
[106] K. Biswas, J. He, I.D. Blum, C.-I. Wu, T.P. Hogan, D.N. Seidman, V.P. Dravid, M.G. Kanatzidis, Nature, 489 (2012) 414-418.
[107] K. Biswas, J. He, Q. Zhang, G. Wang, C. Uher, V.P. Dravid, M.G. Kanatzidis, Nature chemistry, 3 (2011) 160-166.
[108] J.-S. Rhyee, K.H. Lee, S.M. Lee, E. Cho, S.I. Kim, E. Lee, Y.S. Kwon, J.H. Shim, G. Kotliar, Nature, 459 (2009) 965-968.
[109] A.I. Boukai, Y. Bunimovich, J. Tahir-Kheli, J.-K. Yu, W.A. Goddard Iii, J.R. Heath, Nature, 451 (2008) 168-171.
[110] T. Harman, P. Taylor, M. Walsh, B. LaForge, science, 297 (2002) 2229-2232.
[111] R. Venkatasubramanian, E. Siivola, T. Colpitts, B. O'quinn, Nature, 413 (2001) 597-602.
[112] H. Lv, W. Lu, D. Shao, Y. Sun, arXiv preprint arXiv:1404.5171, (2014).

[113] E. Flores, J.R. Ares, A. Castellanos-Gomez, M. Barawi, I.J. Ferrer, C. Sánchez, Applied Physics Letters, 106 (2015) 022102.

[114] S. Chen, K. Cai, W. Zhao, Physica B: Condensed Matter, 407 (2012) 4154-4159.

[115] C. Li, J. Hong, A. May, D. Bansal, S. Chi, T. Hong, G. Ehlers, O. Delaire, Nature Physics, (2015).

[116] L. Zhu, G. Zhang, B. Li, Physical Review B, 90 (2014) 214302.

[117] Z.-Y. Ong, Y. Cai, G. Zhang, Y.-W. Zhang, The Journal of Physical Chemistry C, 118 (2014) 25272-25277.

[118] S. Lee, F. Yang, J. Suh, S. Yang, Y. Lee, G. Li, H.S. Choe, A. Suslu, Y. Chen, C. Ko, Nature communications, 6 (2015).

[119] J. Zhang, H. Liu, L. Cheng, J. Wei, J. Liang, D. Fan, P. Jiang, J. Shi, arXiv preprint arXiv:1601.07302, (2016).

[120] J. Zhang, H. Liu, L. Cheng, J. Wei, J. Liang, D. Fan, P. Jiang, L. Sun, J. Shi, Journal of Materials Chemistry C, (2016).

[121] S. Konabe, T. Yamamoto, Applied Physics Express, 8 (2014) 015202.

[122] H.-H. Fu, D.-D. Wu, L. Gu, M. Wu, R. Wu, Physical Review B, 92 (2015) 045418.

[123] M. Ashino, R. Wiesendanger, A.N. Khlobystov, S. Berber, D. Tománek, Physical review letters, 102 (2009) 195503.

[124] X. Ni, G. Liang, J.-S. Wang, B. Li, Applied Physics Letters, 95 (2009) 192114.

[125] Y. Ouyang, J. Guo, Applied Physics Letters, 94 (2009) 263107.

[126] A. Rodin, L.C. Gomes, A. Carvalho, A.C. Neto, Physical Review B, 93 (2016) 045431.

[127] W. Lu, H. Nan, J. Hong, Y. Chen, C. Zhu, Z. Liang, X. Ma, Z. Ni, C. Jin, Z. Zhang, Nano Research, 7 (2014) 853-859.

[128] A. Castellanos-Gomez, The journal of physical chemistry letters, 6 (2015) 4280-4291.